\documentstyle[psfig]{ewic}

\begin{document}

\header{Spectroscopic Data Analysis for Precise Determination of Stellar Temperatures}

\footer{\emph{Astronomical Data Analysis III}}

\title{Spectroscopic Data Analysis\\
 for Precise Determination
of Stellar Temperatures}

\author{K. Biazzo\\
Universit\`a di Catania - Dipartimento di Fisica e Astronomia\\
via Santa Sofia 78, 95123 Catania, Italy\\
woac.ct.astro.it\\
\email{kbiazzo@ct.astro.it}}
\author{A. Frasca, S. Catalano, E. Marilli\\
INAF - Osservatorio Astrofisico di Catania\\
via Santa Sofia 78, 95123 Catania, Italy\\
woac.ct.astro.it\\
\email{afr@ct.astro.it, scat@ct.astro.it, ema@ct.astro.it}}

\begin{abstract}
The ratio of the depths of spectral lines is a powerful indicator of the effective temperature. The method based on this analysis is capable of discerning small temperature variations of individual stars.

We apply this spectroscopic data analysis to three type of stars, namely an RS~CVn type binary system, a young solar-type star and a Cepheid variable. We show that individual LDRs converted into temperature through calibration relations lead to rotational and pulsational modulation of the average surface temperature with amplitudes of 127~K, 48~K and 1466~K in the three types of stars, with average estimated errors of some tens Kelvin degrees.
\end{abstract}


\section{INTRODUCTION}

Data analysis in astronomy is crucial for the optimal extraction of scientific information. The rapid evolution of technologies related to telescopes and instrumentation needs to be complemented by more and more accurate techniques of astronomical data analysis.

The determination of effective temperature plays a very important role in stellar physics. Indeed, this physical parameter is fundamental for locating stars in the HR diagram, for studies of gravity, mass, abundance and so on. Its measurement is not often as accurate as it is needed. For this reason it is important to find a suitable diagnostics of photospheric temperature and to develop an adequate data analysis software.

The strength of absorption lines has been used for spectral classification since the first beginning of astrophysical spectroscopy, given its dependence on temperature and/or gravity, but it is not directly suitable for accurate temperature measurements. Although it is obvious that stellar spectra vary with temperature, the line strength depends on several parameters, making the interpretation more complicated than one might expect. The depth ratio of two absorption lines (LDR) with different excitation potential proves instead to be a very suitable diagnostics for this purpose, after an appropriate calibration has been done. It actually allows us to measure temperature variations with uncertainties of only a few Kelvin degrees (\cite{Gray94}).

Our goal was to set up a method based onto LDRs to accurately monitor temperature variations that may occur during rotation of magnetically active stars or along the period of a pulsating star. For this reason we used line pairs, in which one line is much more temperature-dependent than the other one or in which the depth variation with temperature has an opposite trend for the two lines. In fact, the ratio of the two line depth parameter is practically independent of rotational and macroturbolence broadening or spectral resolution since both lines are reshaped in the same way. Moreover, small differences in the spectrograph focus or scattered light do not affect LDRs, but, more importantly, differences in line strength coming from differences in chemical abundances largely disappear (for weak lines).

We have performed a calibration of LDRs in absolute temperatures, taking into account the gravity effects, through a set of standard stars. Then, the individual LDRs of our targets have been converted into temperature through the aforementioned calibrations.

The LDR-technique allowed us, at our medium resolution of {\it R} = 14\,000, to measure the modulation of the average surface temperature in two classes of variable stars, i.e. spotted active stars and Cepheid pulsating variables, with sensitivity around 10 Kelvin degrees.

By means of this data analysis, it is possible to derive the fractional spotted area ({\it A$_{\sf rel}$}) and the spot temperature ({\it T$_{\sf sp}$}) in the case of some spotted stars for which there are simultaneous spectroscopic and photometric observations. Moreover, for the study of pulsation properties, this spectroscopic analysis allows us to simultaneously record two fundamental parameters, namely temperature and radial velocity.\\

\section{OBSERVATIONS AND ANALYSIS}
We have acquired the spectra using the REOSC \'echelle spectrograph fed by the 91-cm telescope at Catania Astrophysical Observatory - {\it M. G. Fracastoro} station (Serra La Nave, Mt. Etna). The spectral resolving power was about 14\,000 and the thinned back-illuminated SITe CCD, 1024$\times$1024 pixels, allowed us to record five orders in each frame, spanning from about 5850 to 6700~\AA. The average signal-to-noise ratio ({\it S/N}) at continuum in the spectral region of interest was 200-500 for the standard stars and 100-200 for the target stars.

The data reduction was performed by using the {\sc echelle} task of the IRAF\footnote{IRAF is distributed by the National Optical Astronomy Observatory, which is operated by the Association of the Universities for Research in Astronomy, inc. (AURA) under cooperative agreement with the National Science Foundation.} package following the usual steps: background subtraction, division by a flat field spectrum given by a halogen lamp, wavelength calibration using the emission lines of a thorium-argon lamp, aperture extraction and continuum fitting with a low order polynomial. Additional information about data reductions are given in \cite{Cata02}.

\section{MEASUREMENTS OF LINE-DEPTH RATIO}
The studies of temperature variations on active stars made by Gray and his collaborators are principally based on the lines pair V{\sc{i}} $\lambda$6251.83 \AA-Fe{\sc{i}} $\lambda$6252.57 \AA. Within the spectral range covered by our \'echelle frames there are several pairs of line suitable for temperature determination. The line pairs for each ratio have been chosen to be close one another in order to minimize errors in setting the continuum and in a region of low contamination from telluric lines.

In Table~\ref{tab:lines} we list the wavelengths ($\lambda_{\sf 1}$, $\lambda_{\sf 2}$), the lower excitation potentials (in eV) of the selected lines (\cite{Moore66}) and the working temperature range of each LDR--{\it T$_{\sf eff}$} calibration (see Sect. 4 for description of the calibration). We have considered metal lines because they are the most sensitive to the temperature. Altogether we have identified 30 spectral lines forming 20 pairs suitable for line-depth ratios. These lines were identified through the solar spectrum atlas (\cite{Moore66}) choosing unblended lines. The only exceptions are four lines, each composed of two very close lines with the same temperature dependence, but not resolved at our resolution and behaving as single lines.

\begin{table}[h]
\caption{Spectral lines for developing calibrations using line-depth ratios {\it r = d$_{\lambda_{\sf 1}}$/d$_{\lambda_{\sf 2}}$}, with {\it d} line depth.}
\label{tab:lines}
\begin{center}
\begin{tabular}{ccccccccc}
\noalign{\medskip}
\hline
\hline
\noalign{\medskip}
{\it N} & {$\lambda_{\sf 1}$ }&{  {\it El$_{\sf 1}$} }&{ $\chi$ } & {$\lambda_{\sf 2}$ }&{  {\it El$_{\sf 2}$} }&{ $\chi$ }&{$\Delta${\it T$_{\sf eff}$} }\\
     & (\AA)   &           &  (eV) & (\AA)   &           &  (eV) & (K)\\
\hline
\noalign{\medskip}
1 & 6062.86 & Fe{\sc{i}}~~& 2.18 & 6056.01 & Fe{\sc{i}}~~& 4.73 & 3800--6100\\
2 & 6151.62 & Fe{\sc{i}}~~& 2.18 & 6155.2 & Si{\sc{i}}+Fe{\sc{ii}}~~& 5.62+5.57 & 3900--6000\\
3 & 6180.21 & Fe{\sc{i}}~~& 2.73 & 6155.2 & Si{\sc{i}}+Fe{\sc{ii}}~~& 5.62+5.57 & 3800--6400\\
4 & 6180.21 & Fe{\sc{i}}~~& 2.73 & 6170.4 &  V{\sc{i}}+Fe{\sc{i}}~~& 0.29+4.79 & 3800--6100\\
5& 6199.19 &  V{\sc{i}}~~& 0.29 & 6200.32 & Fe{\sc{i}}~~& 2.61 & 3800--6100\\
6& 6210.67 & Sc{\sc{i}}~~& 0.00 & 6215.2 & Fe{\sc{i}}+Ti{\sc{i}}~~& 4.19+2.69 & 3800--5800\\
7& 6216.36 &  V{\sc{i}}~~& 0.28 & 6215.2 & Fe{\sc{i}}+Ti{\sc{i}}~~& 4.19+2.69 & 3800--6100\\
8& 6243.11 &  V{\sc{i}}~~& 0.30 & 6246.33 & Fe{\sc{i}}~~& 3.60 & 3800--6100\\
9& 6243.11 &  V{\sc{i}}~~& 0.30 & 6247.56 &Fe{\sc{ii}}~~& 3.89 & 3800--6100\\
10& 6246.33 & Fe{\sc{i}}~~& 3.60 & 6247.56 &Fe{\sc{ii}}~~& 3.89 & 3800--6100\\
11& 6251.83 &  V{\sc{i}}~~& 0.29 & 6252.57 & Fe{\sc{i}}~~& 2.40 & 3800--6100\\
12& 6261.1 & Ti{\sc{i}}+V{\sc{i}}~~& 1.43+0.27 & 6176.82 & Ni{\sc{i}}~~& 4.09 & 3800--6100\\
13& 6266.33 &  V{\sc{i}}~~& 0.28 & 6265.14 & Fe{\sc{i}}~~& 2.18 & 3800--6100\\
14& 6268.87 &  V{\sc{i}}~~& 0.30 & 6270.23 & Fe{\sc{i}}~~& 2.86 & 3800--6100\\
15& 6274.66 &  V{\sc{i}}~~& 0.27 & 6270.23 & Fe{\sc{i}}~~& 2.86 & 3800--5800\\
16& 6330.10 & Cr{\sc{i}}~~& 0.94 & 6414.99 & Si{\sc{i}}~~& 5.87 & 3900--6000\\
17& 6355.04 & Fe{\sc{i}}~~& 2.84 & 6419.96 & Fe{\sc{i}}~~& 4.73 & 3900--6100\\
18& 6597.57 & Fe{\sc{i}}~~& 4.79 & 6605.92 & V{\sc{i}}~~& 1.19 & 3800--5800\\
19& 6608.04 & Fe{\sc{i}}~~& 2.28 & 6605.92 &  V{\sc{i}}~~& 1.19 & 3800--5800\\
20& 6608.04 & Fe{\sc{i}}~~& 2.28 & 6597.57 & Fe{\sc{i}}~~& 4.79 & 3800--5800\\
\noalign{\medskip}
\hline
\end{tabular}
\end{center}
\end{table}

\begin{figure}
\begin{center}
\begin{minipage}{14cm}
\centerline{\psfig{file=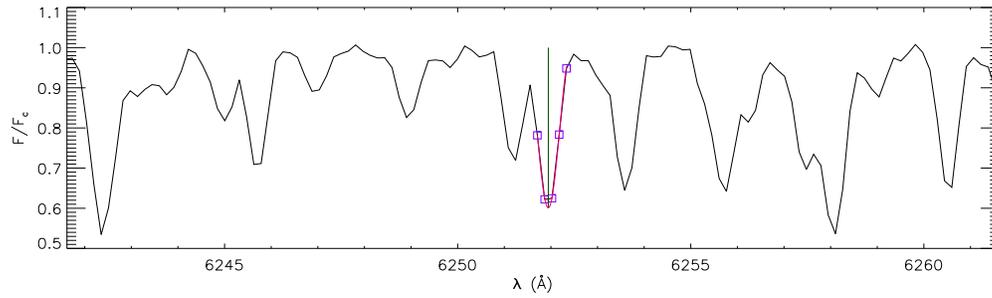,width=14 cm}}
\vspace{-0.5cm}
\caption {Example of normalized spectrum of the standard star $\alpha$~Ari in the 6250 \AA~region. The red line is the polynomial fit on the lowest five points in the core of the Fe{\sc{i}} $\lambda$6252.57 \AA~line. The green segment on this line has the signals in the continuum and in the bottom of the line as extremes.}
\label{fig:norm_spect_alphaAri}
\end{minipage}
\end{center}
\end{figure}

At our resolution, the center of a line profile and therefore the true depth may not concide with an observed data point. So we have developed an IDL ({\it Interactive Data Language}) routine to define the line depth. The lowest five points in the core of each measured line were fitted  with a cubic spline and the minimum of this cubic polynomial was taken as the line depth (see Fig.~\ref{fig:norm_spect_alphaAri}). Writing the line depth {\it d} as
\begin{eqnarray}
d = \frac{S_{\rm c} - S_{\rm b}}{S_{\rm c}} = 1 - \frac{S_{\rm b}}{S_{\rm c}}~,
\label{eq:depth}
\end{eqnarray}
{\noindent where {\it S$_{\sf c}$} and {\it S$_{\sf b}$} are the signals in ADU (Analog to Digital Units) or in photons of continuum and bottom of the line, respectively; the fractional error on {\it d} can be expressed as}
\begin{eqnarray}
\frac{\sigma_{d}}{d} = \frac{\sigma_{\frac{S_{\rm b}}{S_{\rm c}}}}{d} =
\frac{1 - d}{d}~\sqrt{\frac{1}{S_{\rm b}} + \frac{1}{S_{\rm c}}}~.
\label{eq:depth_error}
\end{eqnarray}

\begin{figure}
\begin{center}
\begin{minipage}{11cm}
\centerline{\psfig{file=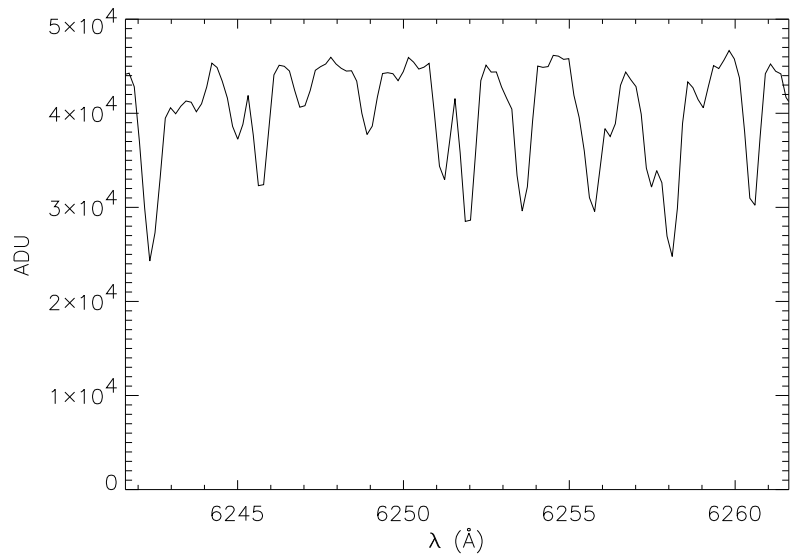,width=11 cm}}
\vspace{-0.5cm}
\caption {Raw spectrum around the 6250 \AA~region of $\alpha$~Ari before the normalization.}
\label{fig:spect_alphaAri}
\end{minipage}
\end{center}
\end{figure}

Assuming that signals in the continuum and in the bottom of the line follow a poissonian statistics, as in the case of a linear detector like a CCD, and considering, as an example, a line depth {\it d} = 0.40 and values of 45000 ADU and 28000 ADU for the continuum and the bottom of the line of a generic raw spectrum (see Fig.~\ref{fig:spect_alphaAri}), given the CCD conversion factor 2.5 e$^-$/ADU we find this typical relative error:

\begin{eqnarray*}
\frac{\sigma_d}{d} \approx 0.72\%~.
\end{eqnarray*}

As we can see below, this can ensure temperature accuracy of 10-20~K.

\section{SPECTROSCOPIC DATA ANALYSIS}
The calibration of LDRs to temperature has been made by means of spectra of 43 standard stars with different spectral type (F6--M4) and luminosity class (V--II) acquired from 2000 to 2002. Since effective temperatures are available for few of our calibration stars, we have adopted effective temperature derived from the empirical relation ({\it B-V})--{\it T$_{\sf eff}$} proposed by \cite{Gray92}. Then, the calibration curves for main sequence (MS) and giant stars have been made separately.

\begin{figure}[t]
 \begin{center}
  \begin{minipage}{11cm}
\vspace{-2cm}
   \psfig{file=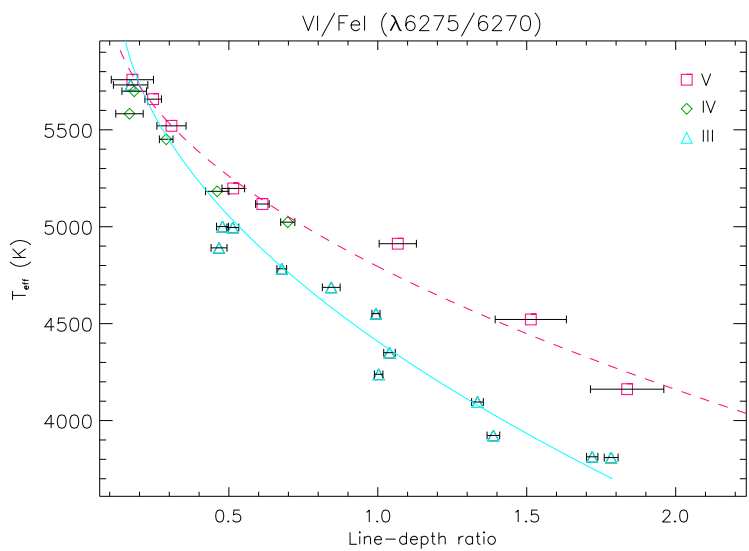,width=11 cm}
\vspace{-7.5cm}
   \caption{An example of effective temperature as a function of LDR. The solid line represents a fit to evolved stars LDR, while the dashed line is the polynomial fit to MS LDR.}
   \label{fig:cal_no_corr}
  \end{minipage}
 \end{center}
\end{figure}
\begin{figure}
 \begin{center}
  \begin{minipage}{10cm}
\vspace{-1cm}
   \psfig{file=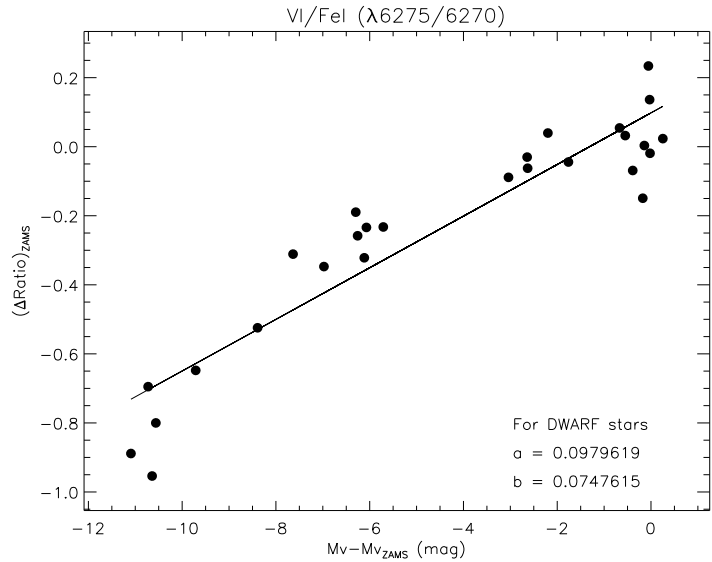,width=10 cm}
\vspace{-0.3cm}
   \caption{Residuals of LDRs compared to the polynomial fit to MS data of Fig.~\ref{fig:cal_no_corr} versus $\Delta${\it M$_{\sf V}$}. The intercept and the slope of the regression line are also reported. An analogous regression line has been obtained for LDRs compared to giant's polynomial fit.}
   \label{fig:resid}
  \end{minipage}
 \end{center}
\end{figure}
\begin{figure}
 \begin{center}
  \begin{minipage}{11cm}
\vspace{-2.5cm}
   \psfig{file=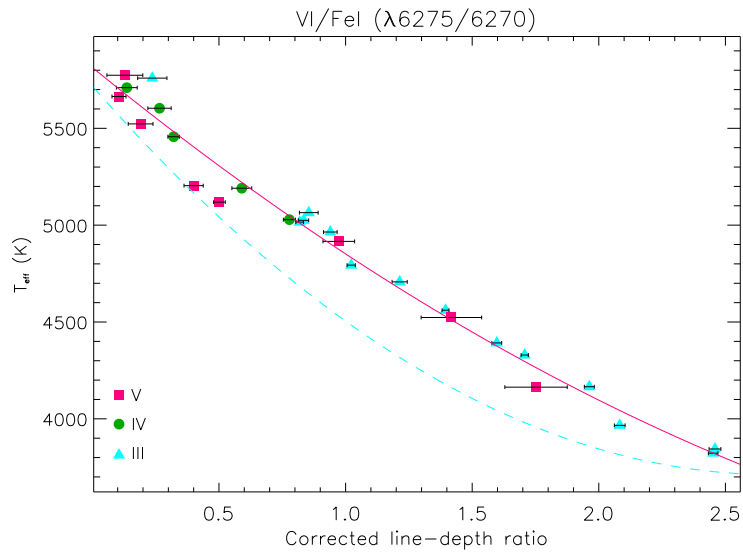,width=11 cm}
\vspace{-7.5cm}
   \caption{Effective temperature as a function of corrected LDR for MS stars. The solid line represents a fit to all data, while the blue dashed line is the polynomial fit to LDR data corrected for giant-star calibration. A similar fit of {\it T$_{\sf eff}$} has been obtained as a function of corrected LDR for giant stars.}
   \label{fig:cal_corr}
  \end{minipage}
 \end{center}
\end{figure}

In Fig.~\ref{fig:cal_no_corr} we show the $\lambda$6275 VI/$\lambda$6270 FeI LDR-{\it T$_{\sf eff}$} calibration as an example. The two different behaviours, for MS and evolved stars, are evident. The gravity dependence of LDRs has been deduced from the residuals of the standard star LDR compared to the fits of Fig.~\ref{fig:cal_no_corr}. These differences correlate very well with the gravity indicator $\Delta${\it M$_{\sf V}$} (Fig.~\ref{fig:resid}), which is the absolute magnitude difference in the HR diagram compared to the ZAMS. The best-fit in Fig.~\ref{fig:resid} has been used to correct the measured LDRs for gravity. A gravity-corrected LDR-{\it T$_{\sf eff}$} calibration for $\lambda$6275 VI/$\lambda$6270 FeI ratio is shown in Fig.~\ref{fig:cal_corr} for MS stars. Data points for stars of different luminosity classes mix well together, allowing a unique correlation, alternatively applicable to MS and giant stars. The temperature sensitivity of each line-depth ratio derived from the slope of the polynomial fit is of the order of 10-20 K for 1\% LDR variation.

\section{LINE-DEPTH RATIO METHOD APPLIED TO THREE TYPES OF STARS}
LDR or temperature variations can occur when hotter or cooler surface features are carried across the disk of an active star as the star rotates or during natural oscillations of a Cepheid star.

On the basis of these considerations we have made observations on three type of stars in order to investigate the applicability of the LDR-analysis.

\subsection{RESULTS ON A RS CVn BINARY}
\begin{figure}  
 \begin{center}
  \begin{minipage}{15cm}
\hspace{-0.1cm}
   \psfig{file=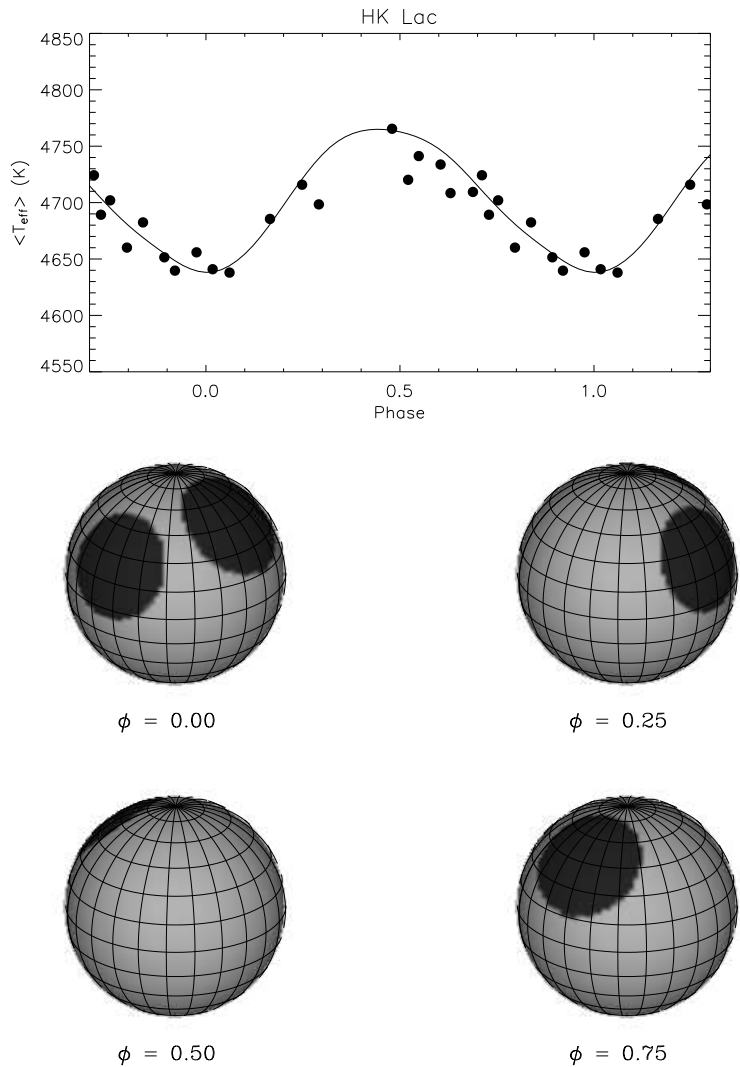,width=15 cm}
\vspace{-0.5cm}
   \caption{Observed (dots) and synthetic (full line) average effective temperature variation of HK~Lac for the two-spots distribution shown in the lower panel at four specific phases. The minimum of temperature curve is around the phase $\phi$ = 0$^{\sf p}$.00 when the two spots are both visible on the disk of the star.}
   \label{fig:hklac_solution}
  \end{minipage}
 \end{center}
\end{figure}

As an example of the application of this analysis technique, Fig.~\ref{fig:hklac_solution} shows the temperature modulation of the slow-rotating single-lined active star HK Lac (HD 209813, K0III) during its rotational period of {\it P$_{\sf rot}$} = 24$^{\sf d}$.4284 (\cite{Gorza71}). The spectra have been acquired from October 2000 to January 2001 and the initial epoch is taken from \cite{Stra93}. The figure displays an average of six LDR-curves converted into temperature curves adopting the calibration for giant stars. The amplitude of the {\it $<T_{\sf eff}>$} curve we measured is about 130~K and the corresponding errors are tipically of a few Kelvin degrees. This temperature variation is the result of the different visibility of cool spots along the rotational period. The upper box of Fig.~\ref{fig:hklac_solution} also showns the synthetic curve, obtained with a two-spots model (lower panel), superimposed on the data points.

\subsection{RESULTS ON A SOLAR-LIKE STAR}
We are also studying the rotational modulation of the temperature through LDR in young solar-type stars. They exhibit intermediate activity levels between that typical of RS CVn binaries and of the Sun.

\begin{figure}
 \begin{center}
  \begin{minipage}{11cm}
   \psfig{file=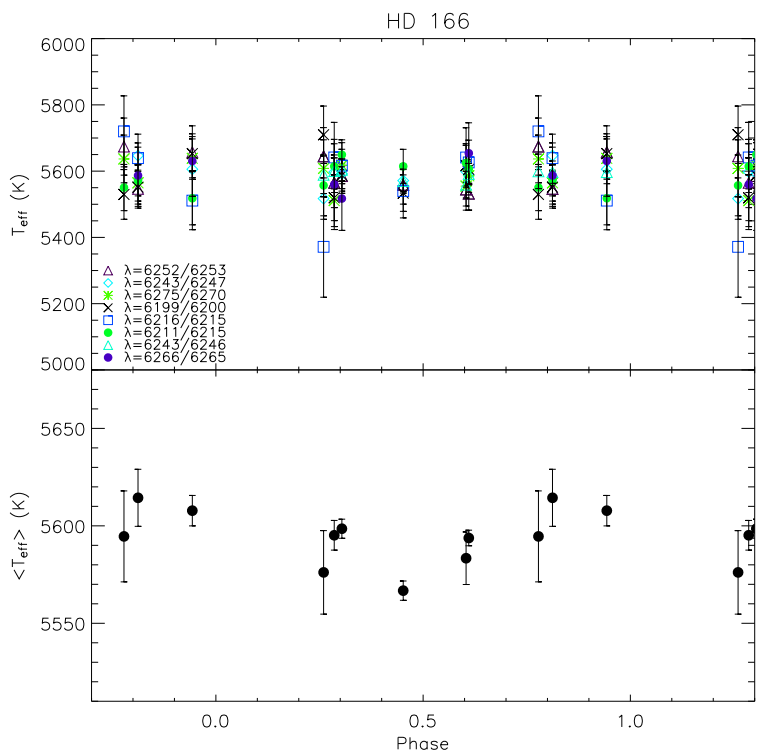,width=11 cm}
\vspace{-1cm}
   \caption{Upper panel: temperature curves of HD 166  obtained from each LDRs. Lower panel: effective temperature as a function of the rotational phase.}
   \label{fig:hd166}
  \end{minipage}
 \end{center}
\end{figure}

Fig.~\ref{fig:hd166} reports preliminary results about the temperature variation of the slow-rotating single main sequence star HD 166 (K0V) during its rotational period of {\it P$_{\sf rot}$} = 6$^{\sf d}$.23\footnote{We have used the mean period and the initial julian day given by \cite{Gaidos00}.}. The observations have been made in 2000 from October 16th to November 16th. HD 166 shows a clear rotational modulation of the average temperature, obtained from 8 LDRs, with an amplitude of about 50~K.

\subsection{RESULTS ON A CEPHEID STAR}
We have also applied the LDR-method to the Cepheid star $\eta$ Aql (F6Ib-G4Ib, {\it P$_{\sf puls}$} = 7$^{\sf d}$.176641\footnote{We have used the period and the epoch of brightest {\it V} given by \cite{Moffett85}.}). The observations have been performed in 2002 between July 22nd and October 6th.

\begin{figure}[t]
 \begin{center}
  \begin{minipage}{10cm}
   \psfig{file=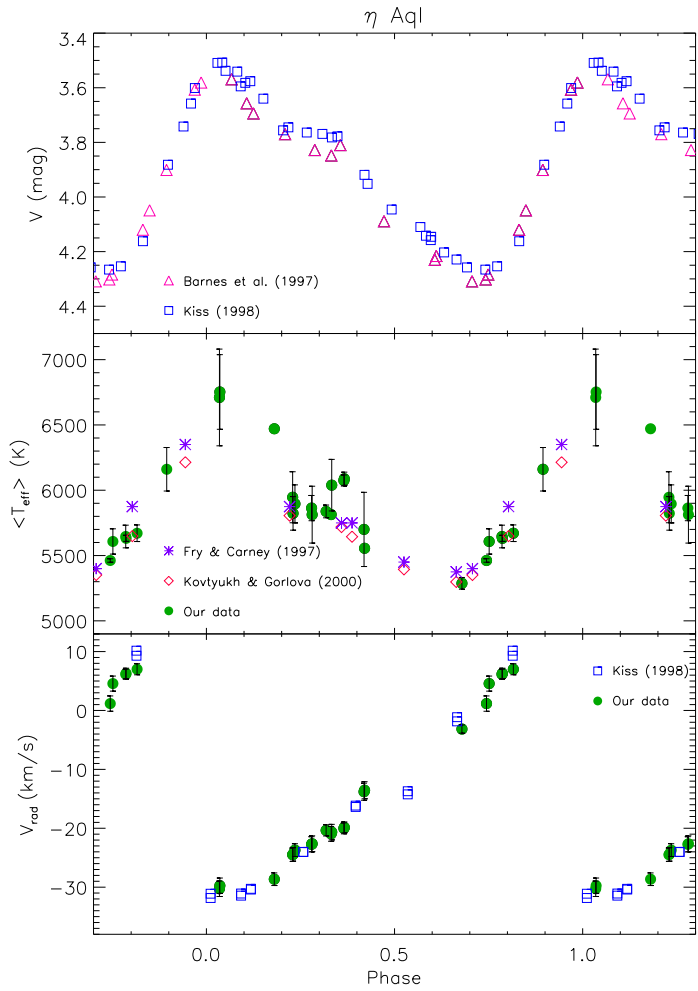,width=10 cm}
   \caption{Upper panel: light curves obtained by \cite{Barnes97} and \cite{Kiss98}. Middle panel: our, \cite{Fry97}'s and \cite{Kovty00}'s average effective temperature as a function of pulsational phase. Lower panel: radial velocity curves from our and \cite{Kiss98}'s observations.}
   \label{fig:etaAql}
  \end{minipage}
 \end{center}
\end{figure}

In Fig.~\ref{fig:etaAql} we plot the light curves obtained by \cite{Barnes97} together with those of \cite{Kiss98}. The difference between the two data sets does not exceed 0.01--0.02 mag, which is within the explicitation in matching different data sets.

In the same figure we show the effective temperature obtained by us (dots) and by \cite{Kovty00} (diamonds) by means of the LDR-method and that derived by \cite{Fry97} (asterisks) using the independence from the excitation potential of Fe{\sc{i}} lines of iron abundance calculated using Kurucz LTE, plane-parallel, atomic and molecular line-blanketing, flux-conserving stellar models. The agreement between these three data sets is good, the difference of values being at the same phase of only few per cent of the total amplitude ({\it $\Delta<T_{\sf eff}>$}$\simeq$ 1500~K).

From the same spectra we derived the simultaneous radial velocity curve, displayed in Fig.~\ref{fig:etaAql} together with that of \cite{Kiss98}. Also in this case, the two sets of points are almost superimposed and the maximum difference compared to our curve amplitude of about 40 km/s is 5\%, i.e. within the typical accuracy of RV measurements.

We can also note that the radial velocity curve of $\eta$ Aql is almost a mirror image of the light curve, so that the luminosity maximum corresponds to the highest velocity of recession. These two curves are lightly shifted, as predicted by the pulsation theory (\cite{Kuka75}).

\section{CONCLUSIONS}
The line-depth ratio method proved to be very effective for measuring the surface temperature in several classes of stars. We have shown that \'echelle spectra, covering a large spectral band and allowing the use of many line pairs, enable us to enhance the precision of effective temperature measurements to few degrees.

Well-defined modulations of the average temperature have been detected in HK~Lac, HD~166 and $\eta$~Aql.

HK~Lac displays a clear rotational modulation of its surface temperature as derived from several LDRs (see \cite{Cata02} for major details). From the combined analysis of contemporaneous temperature and light curves, by applying a spot-model, we can obtain sole solutions for the spot temperature {\it T$_{\sf sp}$} and the relative area coverage {\it A$_{\sf rel}$} with an accuracy of few per cent (\cite{Frasca04}). Spot temperatures we derive are closer to those typical of sunspot penumbra rather than umbra.

For HD~166, our analysis represents a snapshot of its active regions and more observations are required to study any time dependence. It is important to make observations of young solar-type stars because their study provides insight into the behavior of the solar activity and its effect on the planets during the first few hundred million years of solar system history.

We have also derived a strong modulation of $\eta$~Aql temperature during its pulsational phase. Determining accurate values of {\it T$_{\sf eff}$} in Cepheid stars along their pulsational cycle is of paramount importance. For example, precise temperatures provide a reliable localization of variable stars in the Cepheid instability strip, whose width is only about 650~K. Moreover, the combined analysis of temperature, radial velocity and light variations coming from spectroscopic and photometric data allows us to derive accurate values of Cepheid absolute magnitudes. This parameter is of fundamental assistance for distance determination, since the Cepheids are considered the primary standard candles together with the RR~Lyrae variables.

{\noindent {\bf acknowledgements}}
This work has been supported by the Italian {\em Ministero dell'Istruzione, Universit\`a e  Ricerca} (MIUR) and 
by the {\em Regione Sicilia} which are gratefully acknowledged. We also want to thank L. Santagati for the english 
revision of the text.

{}

\end{document}